\documentclass[twocolumn]{revtex4}
\usepackage{amssymb}
\usepackage{amsmath}
\usepackage{graphicx}
\usepackage{epsfig}

\begin{document}

\title{\bf Heat Capacity of Single Crystal Cu$_x$TiSe$_2$ Superconductors}

\author{ J. Ka\v{c}mar\v{c}\'{i}k,$^1$ Z. Pribulov\'{a},$^1$  V. Pal'uchov\'{a},$^1$ P. Szab\'{o},$^1$ P. Husan\'{i}kov\'{a},$^{2,3}$ G. Karapetrov,$^3$  and P. Samuely$^1$}

\affiliation{$^1$ Centre of  Low Temperature Physics at Institute of Experimental Physics, Slovak
Academy of Sciences \& P. J. \v{S}af\'{a}rik University, Park Angelinum 9, SK-04001 Ko\v sice, Slovakia\\
$^2$ Institute of Electrical Engineering, Slovak Academy of Sciences, D\'{u}bravsk\'{a} cesta 9, SK-84104 Bratislava, Slovakia\\
$^3$ Department of Physics, Drexel University, 3141 Chestnut Street, Philadelphia, Pennsylvania 19104, USA}

\date{\today}

\begin{abstract}
We present heat capacity measurements on a series of superconducting Cu$_x$TiSe$_2$ single crystals with different Cu content down to 600 mK and up to 1 T performed by ac microcalorimetry. The samples cover a large portion of the phase diagram from an underdoped to a slightly overdoped region with an increasing superconducting critical temperature and the charge density wave (CDW) order gradually suppressed. The electronic heat capacity as a function of normalized temperature $T/T_c$ shows no difference regardless of the concentration of copper, i.e., regardless of how much the CDW order is developed in the samples. The data analysis reveals consistently a single $s$-wave gap with an intermediate coupling strength $2\Delta/k_BT_c \approx 3.7$ for all samples. 
\end{abstract}

\maketitle


TiSe$_2$ has been one of the most studied systems with charge density wave (CDW) order \cite{CDW1}. The interest has been reinvigorated since a tunable transition from CDW to superconductivity was discovered upon intercalation of TiSe$_2$ by copper \cite{Morosan1} or paladium \cite{Morosan2}. Indeed, an interplay between collective phenomena such as CDW [or spin density waves (SDWs)] and superconductivity is one of the most important issues in modern solid state physics. In Cu$_x$TiSe$_2$ superconductivity appears at $x=0.04$ and culminates at $x=0.08$ with a maximum $T_c=$ 4.15 K, while simultaneously CDW is suppressed with copper intercalation. For dopings close to $x=0.10$, the superconducting critical temperature decreases to 2.5 K.
Kusmartseva $et$ $al.$ \cite{Kusmartseva} discovered that a similar superconducting dome can be induced by high pressures between 2 and 4 GPa applied to undoped TiSe$_2$ showing  a maximum $T_c$ of 1.8 K.

The overall phase diagram, {\it Temperature vs. doping (or pressure)}, is reminiscent of a similar phase diagram of high-$T_c$ cuprates, pnictides, or heavy fermions. 
The question on how the coexistence of strongly correlated states such as CDW and superconductivity affects the superconducting order parameter has been outstanding and might shed light on the mechanism of superconductivity in these systems. 
If, for example, the appearance of the CDW order is connected with a partial gapping of the Fermi surface, it can introduce an anisotropy or even a non trivial symmetry to the superconducting order parameter.   
Superconductivity mediated by density fluctuations, which has been recently predicted, \cite{Lonzarich} can be at play here, particularly in underdoped samples since favorable conditions for such a pairing may be expected on the border of the  CDW transition.  

Several papers addressing the character of the superconducting order parameter have been published on Cu$_x$TiSe$_2$. Li $et$ $al.$ \cite{Li} have found from their thermal conductivity measurements that in Cu$_{0.06}$TiSe$_2$ there is probably only a single superconducting gap which has no nodes but it is finite everywhere on the Fermi surface.  Hillier $et$ $al.$ \cite{Hillier} used muon spectroscopy measurements to obtain information on the temperature dependence of the superfluid density in Cu$_{0.06}$TiSe$_2$ and their data could be fitted accounting for the $s$-wave gap $\Delta$ with a reduced value $2\Delta(0)/k_BT_c = 2.5$, which is much smaller than a BCS canonical value (3.52) for weak superconducting coupling.  Based on their muon spectroscopy mesurements Zaberchik $et$ $al.$ \cite{Zaberchik} observed that while for optimally doped Cu$_x$TiSe$_2$ the superconducting gap has a BCS value, for lower doping where CDW coexists with superconductivity, two-gap superconductivity develops, with one of the gaps being much smaller than the BCS value. Angle-resolved photoemission spectroscopy \cite{Qian} observed that a large electron density of states with a $d$-like character is built with superconducting doping ($x > 0.04$) and that CDW competes with superconductivity in the same band. 

Here, we present comprehensive heat capacity measurements on Cu$_x$TiSe$_2$ superconducting single crystals with four different dopings, from the underdoped to the slightly overdoped regime and superconducting critical temperatures ranging between 2.2 and 3.85 K.
As a result, the electronic heat capacity of all samples can be described by a single $s$-wave gap with a common coupling strength $2\Delta(0)/k_BT_c = 3.7$. The upper critical fields $H_{c2}$ inferred from the heat capacity measurements show the classical BCS-type temperature dependence. The superconducting anisotropy is temperature independent and equal to $\Gamma = 1.8 \pm 0.1$. The angular dependence of $T_c(H)$ in a magnetic field can be fully described by the effective mass model without any deviations that are typical for multigap superconductors. The data strongly suggest a conventional character of superconductivity in the copper doped titanium diselenide, regardless how much the CDW order is developed.


Heat capacity measurements have been performed using an ac technique \cite{Sullivan}. ac calorimetry consists of applying periodically modulated power and measuring the resulting sinusoidal temperature response. In our case, the heat was supplied to the sample at a frequency $\omega\sim$ tens of Hz by a light emitting diode via an optical fiber. The temperature oscillations were recorded by the chromel-constantan thermocouple callibrated in the magnetic field using measurements on ultrapure silicon. Although ac calorimetry is not capable of measuring the absolute values of the heat capacity, it is very sensitive to relative changes in minute samples and it enables continuous measurements. We performed measurements at temperatures down to 0.6 K and in magnetic fields up to 1 T in the $^3$He refrigerator.
Crystals with the dimensions $\sim$ 500 x 500 x 50 $\mu$m$^3$ for samples A and C, and $\sim$ 250 x 250 x 30 $\mu$m$^3$ for samples B and D, were prepared via the iodine gas transport method \cite{Oglesby} with copper intercalation during the crystal growth. The energy dispersive x-ray spectroscopy (EDS) analysis yielded a copper content $x$  $\sim$ 0.086, 0.064, 0.061, and 0.054 for samples A, B, C, and D, respectively.  Sample A is close to optimal doping, while the other samples B, C, and D are from the underdoped region of the phase diagram, temperature vs. copper content.

\begin{figure}[t]
\begin{center}
\resizebox{0.4\textwidth}{!}{\includegraphics{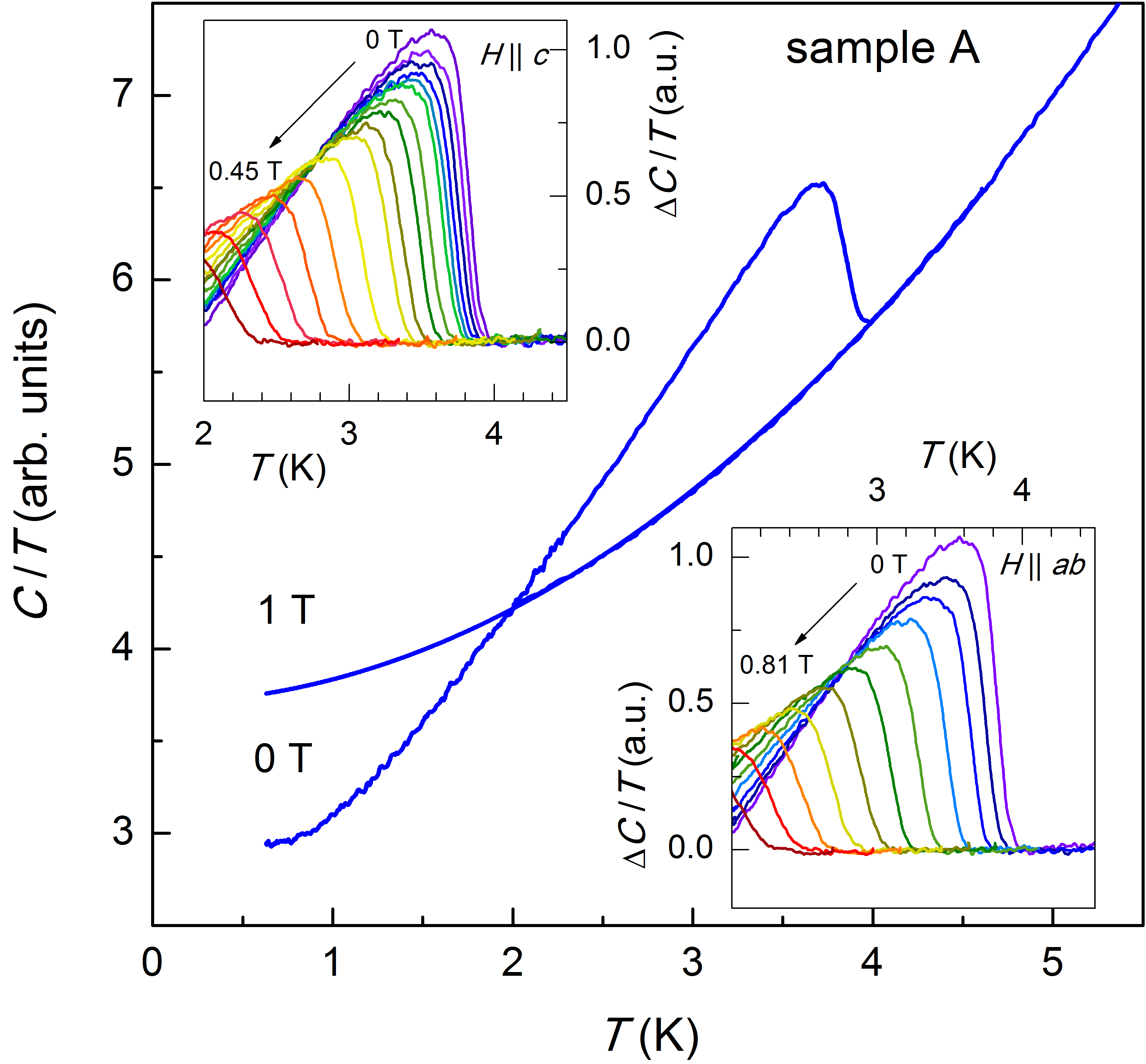}}
\caption{Total heat capacity of  sample A plus addenda in 0 and 1 T. Inset: Temperature dependence of the heat capacity after subtraction of the normal-state background measured in different magnetic fields for a field  in the $c$ direction (upper panel, applied fields 0, 0.01, 0.02, 0.03, 0.04, 0.05, 0.075, 0.1, 0.125, 0.15, 0.2, 0.25, 0.3, 0.35, 0.4, and 0.45 T) and parallel to the $ab$ plane (lower panel, applied fields 0, 0.05, 0.1, 0.18, 0.27, 0.36, 0.45, 0.54, 0.63, 0.72, and 0.81 T).}
\label{fig:fig1}
\end{center}
\end{figure}

\begin{figure}[t]
\begin{center}
\resizebox{0.42\textwidth}{!}{\includegraphics{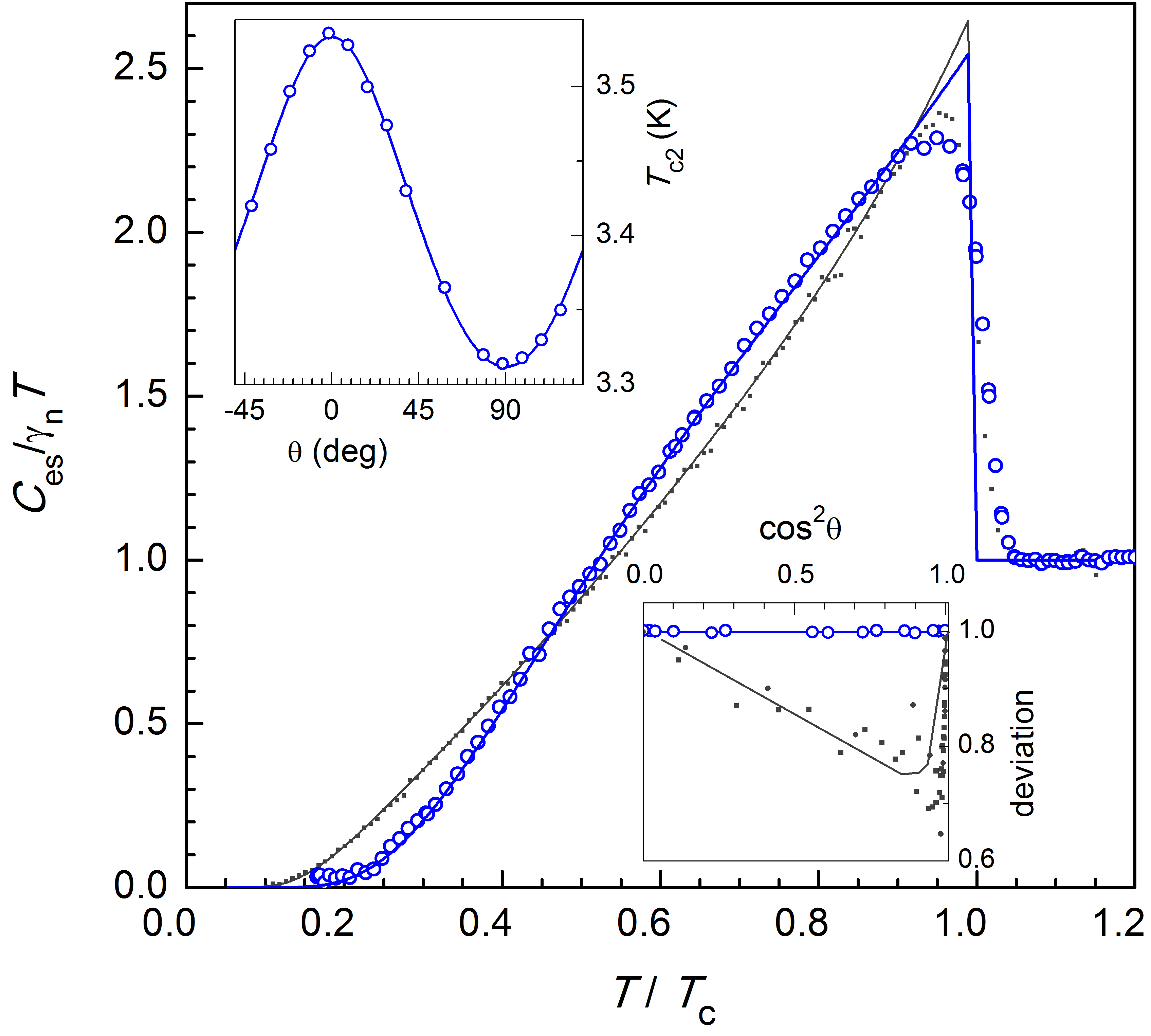}}
\caption{ Superconducting electronic heat capacity of sample A (open symbols) and corresponding single-gap $\alpha$-model fit (thick line), compared to that of NbS$_2$ (Ref. 12) (solid symbols) and $\alpha$-model fit corresponding to the  presence of two energy gaps or one anisotropic gap (thin line). Upper inset: Angular dependence of $T_c$ at 0.15 T (symbols) and a Ginzburgh-Landau fit for a single-gap superconductor (line). Lower inset: Deviation function  ($T_{c2 }$/$T_{c2}^{GL}$)$^2$ for sample A (open symbols) and ($H_{c2}$/$H_{c2}^{GL}$)$^2$ for NbS$_2$ (Ref. 19) (solid symbols); the line is a guide to the eyes.}
\label{fig:fig2}
\end{center}
\end{figure}

Figure 1 summarizes the heat capacity measurements of the sample with the highest copper content (sample A). The main panel plots the total heat capacity of the sample plus addenda in a superconducting and a normal state. In the zero-field measurement  an anomaly at the transition into the superconducting state is clearly visible. It is sharp, indicating the high quality and homogeneity of the crystal. In the 1 T field, superconductivity is suppressed in the whole temperature range and only the normal-state contribution remains. The normal-state heat capacity could be very well fitted with an expression $C$($H$ = 1 T)/$T$=$a+bT^2+cT^4$ corresponding to the electronic and phononic contribution typical for nonmagnetic metal. We used this dependence to extract the electronic heat capacity from our measurements. The heat capacity of the lattice is the same in the superconducting as well as in hte normal state, and the magnetic-field dependence of the addenda is negligible. Thus by subtracting the normal-state measurement from the one in the superconducting state we eliminate contributions from the phonons and from the addenda and what remains is a temperature dependence $\Delta C/T$ = $C(H$ = 0$ T)/T$ - $C(H$ = 1$ T)/T$ = $C_{es}/T$ - $ C_{en}/T$, where $C_{es}$ and $C_{en}$ is the electronic heat capacity of the sample in the superconducting and normal state, respectively. The only assumption in this procedure is the absence of the magnetic-field dependence of the addenda. This has been previously verified in numerous experiments using the same thermocouple wires (see, e.g., Refs. 12 and 13) and is also confirmed here by the entropy conservation rule (there is no difference in entropy above $T_c$ when integrating the $C/T$ curve in 0 T or in 1 T),  proving the thermodynamic consistency of our measurements. The resulting temperature dependence of $\Delta C/T$ is plotted in both insets of Fig. 1 as the rightmost curve. The critical temperature of the superconducting transition in zero field was determined from the local entropy balance around the anomaly, giving $T_c$ = 3.85 K. Similarly one can obtain the critical temperature in different magnetic fields. The insets of Fig. 1 present the results for two principal field orientations, parallel with basal ($ab$) planes and perpendicular to them. In both cases the anomaly at the transition is gradually shifted to lower temperatures with increasing field. Despite some broadening at higher fields the anomaly remains very well resolved at all fields.

The temperature dependence of $\Delta C$/$T$ in zero field was inspected in detail. The difference in entropy between the superconducting and normal state has been calculated as $\Delta S(T^\prime)$ = $\int_{T^\prime}^{T_c}{(\Delta C/T})dT$. From the second integration of the data we obtain a temperature dependence of the thermodynamic critical field $H_c$ as $H_{c}^2$($T^{\prime\prime}$) = $8\pi$$\int_{T^{\prime\prime}}^{T_c}{\Delta S(T^\prime)}dT^\prime$. Since the results of ac calorimetry measurements are in arbitrary units, such a calculated $H_c$ is also in arbitraty units. Still it bears information about the coupling strength in the system. The ratio [$T/H_c$(0)](d$H_c/dT)|_{T\rightarrow T_c}$  is equal to $\Delta(0)/k_BT_c$ \cite{Toxen}. Taking the value of $H_c$(0) = 1.71 and the derivative of $H_c$ in the vicinity of $T_c$ equal to 0.82, we get the coupling ratio $2\Delta/k_BT_c$ = 3.7.

To estimate the coupling strength of the superconducting electrons we compared the electronic heat capacity $C_{es}$/$\gamma_n$$T$ = $\Delta C$/$\gamma_n$$T$ + 1, where $\gamma_n$ = $C(H$ = 1 T)/$T_{|T\sim0K}-C(0$ T)/$T_{|T\sim0K}$ with the so-called $\alpha$ model \cite{Padamsee} based on the BCS theory. The only  parameter in this model is the gap ratio 2$\Delta/kT_c$. The model may be also adjusted to account for two-gap superconductivity [see, for example the case of MgB$_2$ (Ref. 16) or NbS$_2$ (Ref. 12)) or an anisotropic energy gap in the system if necessary. Figure 2 shows the electronic heat capacity $C_{es}$/$\gamma_n$$T$ of sample A in a normalized scale (open circles) and the corresponding single-gap $\alpha$-model fit (thick line) with 2$\Delta/k_BT_c$ = 3.7. The fit reproduces the jump at $T_c$ and also the overall temperature dependence of the electronic heat capacity in very good agreement. For illustration the electronic heat capacity of NbS$_2$ measured down to 0.6 K with a corresponding $\alpha$-model fit from our previous work \cite{Kacmarcik} is shown as well. In that case we showed that the curve is best fitted with the two-gap model with small and large gaps 2$\Delta_S/k_BT_c$ = 2.1, 2$\Delta_L/k_BT_c$ = 4.6, respectively and their relative contributions $\gamma_{S}/\gamma_{n}$ = 0.4. The model with one anisotropic energy gap can describe the NbS$_2$ data as well.  

Comparing the heat capacity in these two dichalcogenides, we can see that while the jump at the anomaly is comparable in both cases, there are significant differences in the overall temperature dependence, mainly in the low temperature region. In the case of  NbS$_2$, due to the small gap $\Delta_s$, the electronic heat capacity starts to increase from zero at much lower temperatures. It is only for significantly higher temperatures that the thermal energy becomes sufficient for the excitation of quasiparticles across the gap of Cu$_{0.085}$TiSe$_2$ and the heat capacity starts to increase as well. This is consistent with a single, much higher energy gap value.  

On sample A we have also performed the heat capacity measurements at a fixed field oriented at different angles with respect to the $ab$ plane. The upper inset of Fig. 2 shows the angular dependence of the transition temperature $T_{c2}(\theta)$ at 0.15 T. The anisotropy of the BCS (single $s$-wave gap) superconductor is described by the effective mass model within the Ginzburg-Landau theory, where $T_{c2}(\theta)=T_{c0}+H\sqrt{cos^2(\theta)+\Gamma^2sin^2(\theta)}/(\partial H^{ab}_{c2}/\partial T)$, where $T_{c0}$ is the zero-field transition temperature. As can be seen in the inset, this formula describes our data perfectly. Quantitatively it is documented in the lower inset where  a deviation function $(T_{c2}/T_{c2}^{GL})^2$ is plotted by the open symbols, showing no difference between the data and the theory. This is very different from similar measurements on different types of multigap superconductors such as MgB$_2$ \cite{Rydh}, iron pnictides, \cite{Welp} or NbS$_2$ \cite{Pribulova}. In all those cases the deviation function reveals a typical shape, shown in the lower inset by the solid symbols where our previous measurements on NbS$_2$ are presented \cite{Pribulova}. Here the deviation function was calculated from heat capacity and magnetization measurements as $(H_{c2}/H_{c2}^{GL})^2$. This is another strong argument supporting the presence of only a single gap in our Cu$_x$TiSe$_2$ sample.

Similar comprehensive measurements and data treatment were performed on all the studied samples with different copper content. However, due to limited space, we present only a summary of the main results. 

\begin{figure}[t]
\begin{center}
\resizebox{0.45\textwidth}{!}{\includegraphics{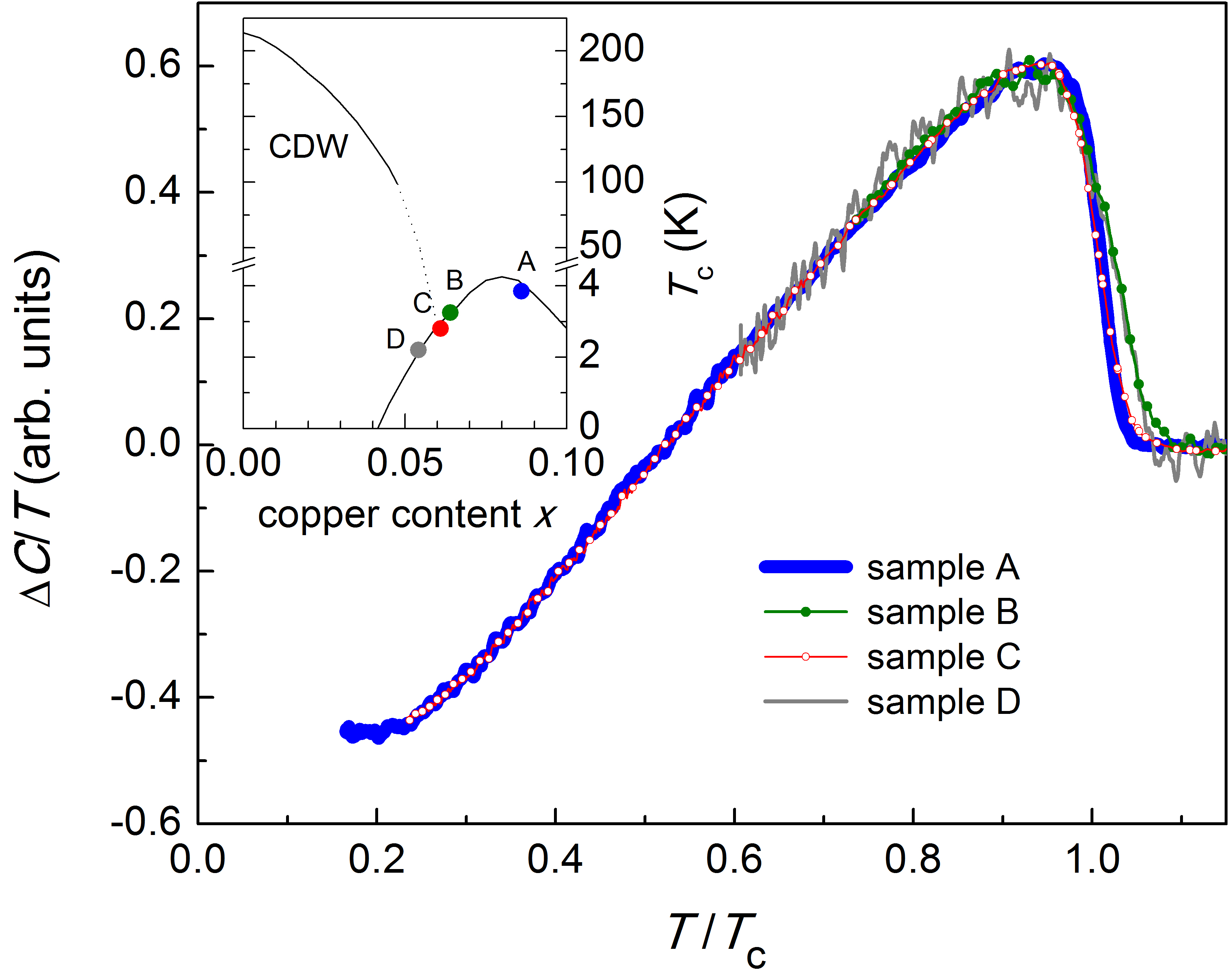}}
\caption{The temperature dependence $\Delta C/T$ of samples A, B, C, and D in a temperature scale normalized to the $T_c$ of each sample. Inset: The phase diagram as proposed by Morosan $et$ $al.$ (Ref. 2) and the critical temperature of our samples with respect to their copper content (large symbols).}
\label{fig:fig3}
\end{center}
\end{figure}

The superconducting critical temperature of each sample was determined from the local entropy balance around the anomaly in $\Delta C$/$T$ as described before, giving the values of $T_c$ $\sim$ 3.85, 3.25, 2.8, and 2.2 K for samples A, B, C, and D, respectively. Taking the amount of copper content from EDS measurements we can construct the $T_c$-$x$ phase diagram as shown in the inset of Fig. 3. The dome-shaped doping dependence of $T_c$ as suggested by Morosan $et$ $al.$ \cite{Morosan1} in polycrystalline material is well reproduced by our single crystals. Sample A is overdoped, far from CDW order. Samples B, C, and D are from the underdoped region with $T_c$'s suppressed with decreasing $x$ and thus are gradually  immersed in the region with the CDW phase more pronounced. This fact has also been evidenced by direct scanning tunneling microscopy (STM) images of charge density waves on the samples from the same batch by Iavarone $et$ $al.$ \cite{Iavarone}. There, CDW patterns with the lowest intensity first appear in the sample Cu$_{0.06}$TiSe$_2$, and in samples deeper in the underdoped regime the amplitude of the charge modulation increases. 

Figure 3 presents the main outcome of this Rapid Communication. It plots the electronic heat capacity of all samples in the temperature scale normalized to their $T_c$. Due to their arbitrary units the curves have been rescaled on top of each other by a corresponding factor to have the same jump at $T_c$ (or the same value at the peak). Surprisingly, in this normalized scale all curves overlap without any significant differences in the overall temperature dependence. This is in contrast to what was proposed by Zaberchik $et$ $al.$ \cite{Zaberchik}  and it clearly shows that superconductivity at all levels of dopings (from slightly overdoped to deeply underdoped) can by described by a single $s$-wave superconducting gap that scales with $T_c$. The coupling strength remains the same for all dopings. Moreover, the angular dependence of $T_{c}(H,\theta)$ has been determined also on sample C and, similarly to what was presented for sample A, it shows no deviation from Ginzburg-Landau theory for a single-gap superconductor. 

\begin{figure}[t]
\begin{center}
\resizebox{0.45\textwidth}{!}{\includegraphics{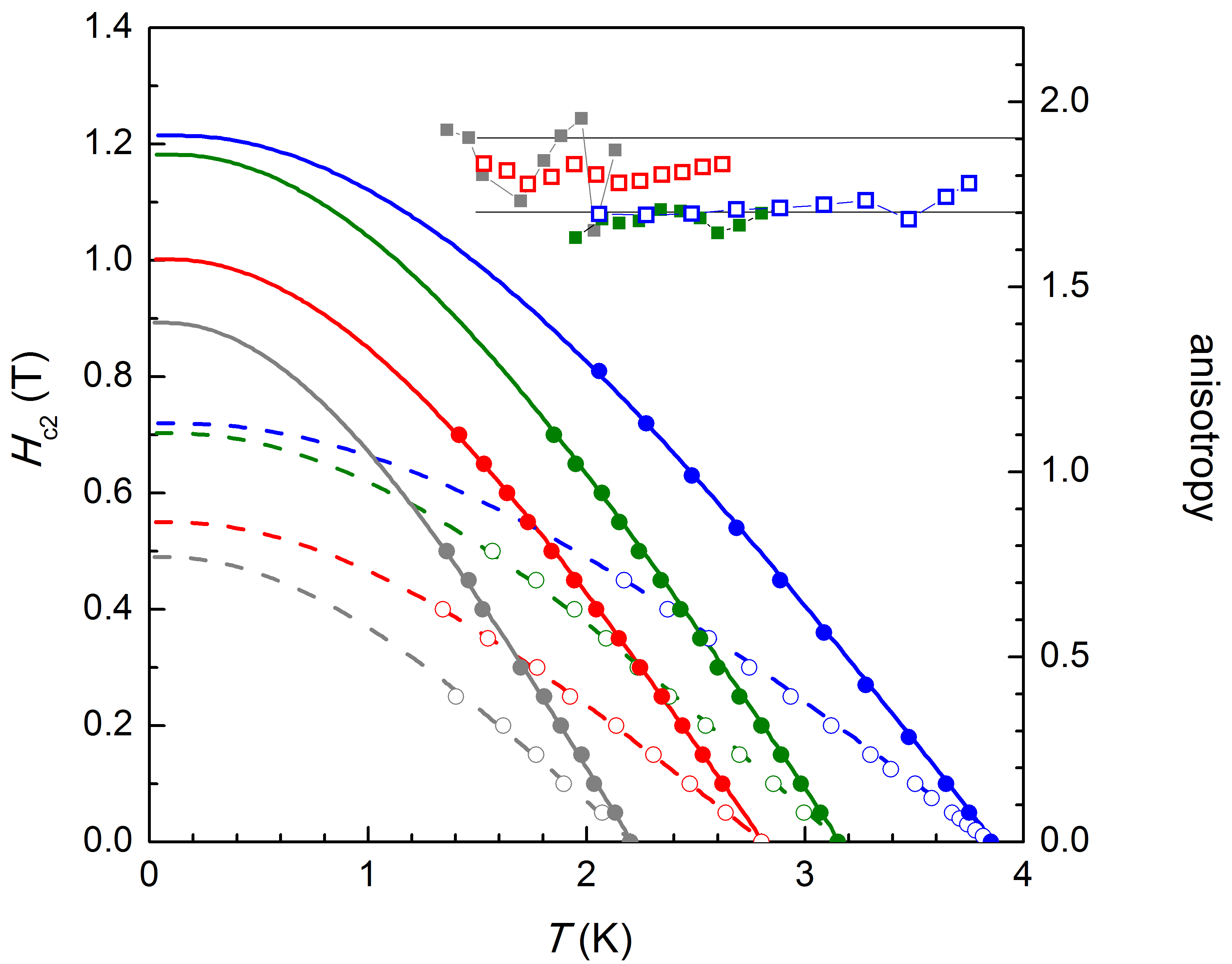}}
\caption{Circles: Upper critical field of the samples for field orientation parallel  (solid circles) and perpendicular to the $ab$ planes (open circles) and corresponding theoretical curves from the WHH model (lines). Squares: Superconducting anisotropy $\Gamma$=$H_{c2}^{ab}$/$H_{c2}^c$; the right $y$ axis applies.}
\label{fig:fig4}
\end{center}
\end{figure}

Figure 4 summarizes the upper critical field of the studied samples. For each sample it was derived from $\Delta$$C$/$T$ in specific magnetic fields for the field directed in the two main crystallographic orientations of the sample - paralel (solid symbols) and perpendicular (open symbols) to the $ab$ planes. The local entropy balance around the anomaly has been taken as a criterion to determine $H_{c2}$ for each magnetic-field measurement. The temperature dependence of $H_{c2}$ reveals a linear behavior close to the critical temperature and a gradual deviation from linearity at lower temperatures. Though measured only in a limited temperature range, it can be described  in the framework of the Werthamer-Helfand-Hohenberg (WHH) theory \cite{WHH}. Lines represent the respective fit for each $H_{c2}$ curve. The values for sample A are close to those published previously by Husanikova $et$ $al.$, \cite{Husanikova} determined from measurements of magnetoresistance on sample Cu$_{0.1}$TiSe$_2$. In Fig. 4 the superconducting anisotropy defined as $\Gamma$ = $H_{c2}^{ab}$/$H_{c2}^{c}$ for every sample is plotted as well. It is independent on temperature with the value between 1.7 and 1.9. These values are in agreement with the study of Morosan $et$ $al.$ \cite{Morosan3} on a slightly underdoped sample giving the anisotropy value 1.7. It is worth noticing that multigap superconductors such as, e.g. MgB$_2$, show $\Gamma$ temperature dependent \cite{Lyard}.

To conclude, heat capacity was measured on a series of superconducting Cu$_x$TiSe$_2$ single crystals with different copper dopings from the underdoped to the overdoped region. The temperature dependence of the electronic heat capacity can be described by the unique BCS formula with a single $s$-wave gap of intermediate coupling strength $2\Delta/k_BT_c \approx 3.7$ for all samples down to the underdoped regime where CDW order coexists with superconductivity.  Neither the angular dependence of the upper critical field (critical temperature) nor the superconducting anisotropy show any indications of an unconventional or multiple order parameter.
 

We acknowledge the assistance of V. Komanick\'y with the EDS measurements.
This work has been supported by the following projects:
CFNT MVEP - Centre of Excellence of Slovak Academy of Sciences, 
FP7  MNT ERA.Net II. ESO,  
EU ERDF grant No. ITMS26220120005, 
Slovak Research and Development Agency Contracts No. APVV-0036-11, VVCE-0058-07, and VEGA No. 2/0135/13. 
Liquid nitrogen has been sponsored by the U.S. Steel Ko\v sice.

\end{document}